\documentclass{article}

\usepackage{microtype}
\usepackage{graphicx}
\usepackage{subfigure}
\usepackage{amsmath,amssymb,amsfonts}
\usepackage{algorithm}
\usepackage{algorithmic}
\usepackage{booktabs} % for professional tables

\usepackage{hyperref}

\usepackage[accepted]{icml2021}

\icmltitlerunning{Saliency Guided Adversarial Training for Learning Generalizable Features}

\begin{document}

\twocolumn[
\icmltitle{Saliency Guided Adversarial Training for Learning Generalizable Features with Applications to Medical Imaging Classification System}



\icmlsetsymbol{equal}{*}

\begin{icmlauthorlist}
\icmlauthor{Xin Li}{to}
\icmlauthor{Yao Qiang}{to}
\icmlauthor{Chengyin Li}{to}
\icmlauthor{Sijia Liu}{goo}
\icmlauthor{Dongxiao Zhu}{to}

\end{icmlauthorlist}

\icmlaffiliation{to}{Wayne State University, Detroit, USA}
\icmlaffiliation{goo}{Michigan State University, East Lansing, USA}

\icmlcorrespondingauthor{Dongxiao Zhu}{dzhu@wayne.edu}

\icmlkeywords{Machine Learning, ICML}

\vskip 0.3in
]
\printAffiliationsAndNotice{}




\begin{abstract}
This work tackles a central machine learning problem of performance degradation on out-of-distribution (OOD) test sets. The problem is particularly salient in medical imaging based diagnosis system that appears to be accurate but fails when tested in new hospitals/datasets. Recent studies indicate the system might learn shortcut and non-relevant features instead of generalizable features, so-called `good features'. We hypothesize that adversarial training can eliminate shortcut features whereas saliency guided training can filter out non-relevant features; both are nuisance features accounting for the performance degradation on OOD test sets. With that, we formulate a novel model training scheme for the deep neural network to learn good features for classification and/or detection tasks ensuring a consistent generalization performance on OOD test sets. The experimental results qualitatively and quantitatively demonstrate the superior performance of our method using the benchmark CXR image data sets on classification tasks.  
\end{abstract}

\section{Introduction}
\label{submission}
Learning good feature representation that generalizes well to Out-Of-Distribution (OOD) test sets is a central challenge in machine learning. Recently, Deep Neural Network (DNN) has demonstrated impressive performance in classification and objection detection tasks on Independent and Identically Distributed (IID) test sets \cite{li2020improving}. Model regularization techniques, e.g., those based on parameter sparsity and loss function smoothing, used in conjunction with adversarial training, have been proven effective on mitigating {\it robust overfitting} \cite{rice2020overfitting} on IID test sets. Nevertheless, the performance degradation on OOD test sets remains a salient problem \cite{shao2020open}. One observation is that the current approach introduces a nearly ideal scenario for DNN to learn spurious shortcuts or non-relevant features \cite{geirhos2020shortcut} that do not exist in OOD test sets. In medical imaging systems, the problem becomes even more salient due to the significant distribution shift between imaging data sets acquired from different hospitals, populations, and time periods. As a result, the AI imaging system that is seemingly effective on training sets often does not generalize well to new hospitals or data sets \cite{degrave2021ai}. Fortunately, in the relatively closed medical imaging environment, we are not so much concerned about adversarial OOD test sets. Instead, we consider how to leverage adversarial IID data sets for learning good features. 

%Another observation is that the more recent open-set recognition methods for OOD test sets can successfully identify OOD samples but fail on adversarial examples and adversarial defense methods are not successful in identifying adversarial OOD samples as well \cite{shao2020open}.  
\begin{figure}[ht]
	\centering
	\includegraphics[width=0.48\textwidth]{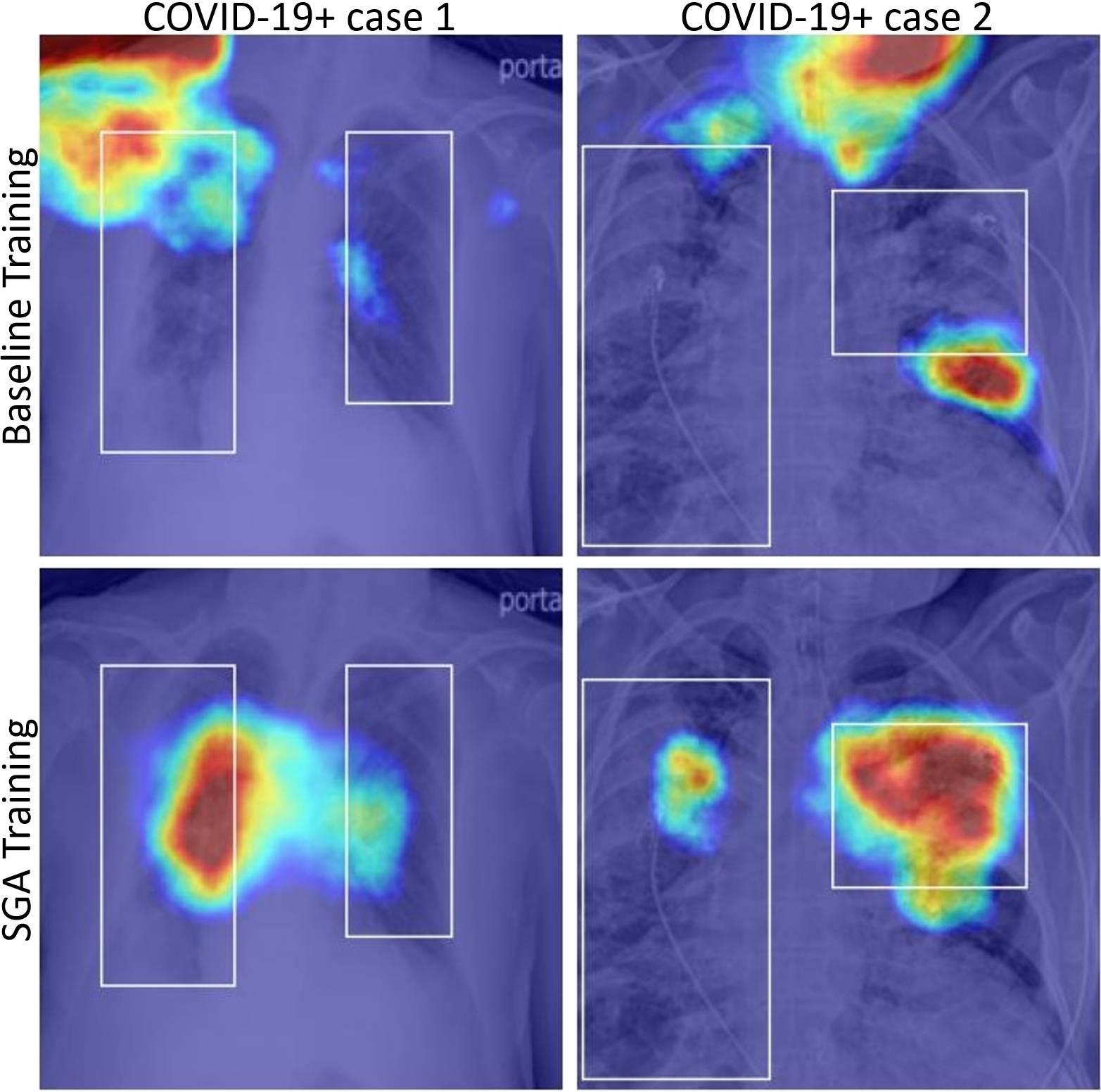}
	\caption{Motivation examples to illustrate the shortcut features (top left) and non-relevant features (top right). Good features are highlighted with high salience in the second rows, overlapping with radiologists' annotations. The heatmap based DNN interpretations are generated by FullGrad.}
	\label{fig:feature1}
\end{figure}

Here we give two motivation examples to illustrate the above-mentioned problem using salience based model explanation methods. Salience methods are a main body of explainable machine learning approaches that quantify individual attribution of input features to the output. Exemplar methods include Integrated Gradient \cite{sundararajan2017axiomatic,pan2021explaining}, Grad-CAM \cite{selvaraju2017grad}, Saliency Map \cite{simonyan2013deep} and FullGrad \cite{srinivas2019full}. In Figure \ref{fig:feature1}, the left example shows the shortcut features and the right example shows non-relevant features are used to predict COVID positive cases. Both types of features can harm the generalizability of the DNN models to OOD test sets. The bounding box represents radiologists' annotated pathological features also known as symptom reasoning \cite{yang2019causal}, which align well with the high saliency regions highlighted by our saliency guided adversarial training (SGA) scheme, but not the baseline cross-entropy based training.     
  
How can we develop an effective robust training scheme to learn the good features for generalizing to test sets? There are four types of test cases, i.e., IID, Adversarial IID, OOD, and Adversarial OOD. Adversarial test cases are rare in the medical imaging system since it is a relatively closed environment that takes pre-processed clean imaging inputs. Thus a major challenge is that the AI system tends to learn and exploit non-relevant features and/or shortcut features, as opposed to generalizable features from training, leading to downgraded performance on OOD test sets. 

Recent studies \cite{maguolo2021critic,cohen2020limits} demonstrate that CXR classification systems might depend more on nuisance features generated by different medical devices with various manufacturing standards and acquisition parameters. Similar to adversarial perturbation, those nuisance features do not impede human recognition but is obvious to DNN models, particularly when they lay on extremely clean background around the CXR borders \cite{li2020robust}. As shown by case 1 (top left in Figure \ref{fig:feature1}), model using those shortcut features would have a poor generalization on OOD test sets. To enhance the OOD generalization, \citet{yi2021improved} proves that model trained robust to adversarial perturbation generalizes well on OOD data. Base on their work, we further hypothesize that adversarial training \cite{madry2017towards} can eliminate those shortcut features since adversarial perturbation are also imperceptible and usually considered as the worst case noise. On the other hand, \citet{ismail2021improving} demonstrate a saliency guided training encourage the model to learn and assign low gradient values to non-relevant features in model predictions, resulting in a more faithful learning of the intended features. Both have been developed to improve feature learning for better generalizability. Here we propose a novel saliency guided adversarial training for better feature representation learning. The saliency guided component eliminates the non-relevant features by reducing their gradient values, whereas adversarial training enhances the robustness of model against learning shortcut features by adding noise to the most relevant features. Using CXR based experiments, we demonstrate that our SGA training scheme learns generalizable features for improving the test performance on the OOD CXR data sets.

% We hypothesize that adversarial training \cite{madry2017towards} can eliminate non-relevant features whereas saliency guided training \cite{ismail2021improving} can ensure the model's performance on OOD test sets via removing shortcut features. Both have been developed to improve feature learning for better generalizability. Here we propose a novel saliency guided adversarial training to eliminate the non-relevant features and suppress shortcut features. Using CXR based experiments, we demonstrate that our SGA training scheme learns generalizable features for improving the test performance on the OOD CXR data sets.
 
%\begin{table}
%	\small{
%		\caption{\small{Adversarial ID-OOD Test Sets. $L_c^{\ast}$ represents a discriminative loss function and $L_d^{\ast}$ a generative likelihood function. * represents adversarial setting.}}\label{tab:ID-OOD}
%		\vspace{0.1in}
%		\begin{tabular}{lll}
%			Input Types	& IID Test  & OOD Test  \\\hline
%			Benign &  $\min_{W} L_c (.)$ +++        & $\min_{\Phi} L_d (.)$ +         \\\hline
%			Adversarial     & $\min_{W^{\ast}} L_c^{\ast}(.)$    +     & $\min_{W^{\ast}, \Phi^{\ast}} L_c^{\ast}(.) + \lambda L_d^{\ast}(.) $    +     \\\hline
%			\vspace{-20pt}
%	\end{tabular}}\label{tab:testsets}
%\end{table}

%(\textbf{Xin, please insert the figure illustrating the performance degradation the real-world test sets})\\

\section{Related Work}
To enhance model robustness against adversarial ID and OOD examples, various robust training techniques have been proposed, including those training with augmented adversarial examples, aka, adversarial training \cite{madry2017towards}, robust regularization \cite{tack2021consistency,chen2019robust,boopathy2020proper}, and improved loss functions, e.g., \cite{li2020improving}. Here we focus on the related works in robust training.  
\subsection{Robust training for IID test cases}
One line of approaches \cite{kurakin2016adversarial,sinha2017certifiable,zhang2019you,shafahi2019adversarial} are based on adversarial training \cite{goodfellow2014explaining} and achieve effective robustness against different adversarial attacks, where the training dataset is augmented with adversarial examples. Adversarial training has also been shown effective in learning robust features for enhanced robustness \cite{ilyas2019adversarial,madry2017towards}. However, these methods have trade-offs between accuracy and adversarial robustness \cite{tsipras2018robustness} and are computationally expensive in adversarial sample generation \cite{zhang2019you}. To reduce the computational burden, Shafahi et al. \cite{shafahi2019adversarial} propose a training algorithm, which improves the efficiency of adversarial training by updating both model parameters and image perturbation in one backward pass. \cite{wong2020fast} discover that it is possible to train empirically robust models using a much weaker and cheaper FGSM based adversary training combined with random initialization. 

Another line of defending strategy against adversaries, other than augmenting the training dataset, is to learn robust feature representations by using model ensembles or altering network architectures \cite{taghanaki2019kernelized,mustafa2019adversarial,tramer2017ensemble,liao2018defense,pang2019improving,xu2017feature,meng2017magnet}. For example, \cite{taghanaki2019kernelized} augment DNNs with the radial basis function kernel to further transform features via kernel trick to improve the class separability in feature space and reduce the effect of perturbation. \cite{mustafa2019adversarial} propose a prototype objective function, together with multi-level deep supervision. Their method ensures the separation in feature space between classes and shows significant improvement of robustness. \cite{pang2019improving} develop a strong ensemble defense strategy by introducing a new regularizer to encourage diversity among models within the ensemble system, which encourage the feature representation from the same class to be close. Although these approaches avoid the high computational cost of adversarial training, they have to modify the network architecture or require an extra training process, limiting the flexibility in adapting to different tasks.

%More efficient approaches can be designing new loss functions to improve model adversarial robustness. By explicitly imposing regularization on latent features, DNNs are encouraged to learn feature representation with more inter-class separability and intra-class compactness \cite{pang2019rethinking,elsayed2018large,mustafa2019adversarial}. For example, \cite{pang2019rethinking} propose a Max-Mahalanobis center (MMC) loss to learn discriminative features. They first calculate Max-Mahalanobis \cite{pang2018max} centers for each class and then encourage the features to gather around the centers using Center Loss \cite{wen2016discriminative}. However, the assumption of geometrical compactness for latent features (in terms of Euclidean distance or $L_2$-norm) may not hold due to inherent intra-class variations in the data and usually requires suitable assumptions on distribution of the latent features. Differently, another line of work, e.g., PC loss with logit constraints \cite{li2020improving}, and GCE loss \cite{chen2019complement}, avoid this issue by learning probabilistically compact features without geometric assumptions. The former enlarges a gap of probabilities between the true class and the first several most probable false classes whereas the later encourages the predicted probabilities of the false classes to be equally distributed.

\subsection{Robust training for OOD test cases} 
When trained on IID examples, DNNs are known to fail against test inputs that lie far away from training distribution, commonly referred to as OOD examples \cite{hendrycks2016baseline}. Recent robust training for detecting ODD test cases considers a multi-class dataset as IID (e.g., CIFAR-10) and uses examples from another multi-class dataset as OOD (CIFAR-100) \cite{liang2017enhancing,hendrycks2016baseline,wei2020minimum,lee2017training}. Existing works either train an OOD detector and a classifier sequentially \cite{sehwag2019analyzing,li2020defending} or simultaneously \cite{anonymous2021informative}. For example, \cite{sehwag2019analyzing} employ adversarial training on IID data as well as OOD examples that are close to IID examples to improve learning robust features. These approaches work well for the so-called closed-world detection where OOD examples are either with simpler data modalities (e.g., medical images with large shared backgrounds) or closer to IID examples (CIFAR-10 versus CIFAR-100). Different from IID detection tasks where robust discriminative features are learned from labeled training data, OOD detection needs to learn {\it high-level}, {\it task-agnostic} and {\it semantic} features from the IID dataset to detect diverse OOD inputs at the test time. 

More recent OOD detection approaches are self-supervised representation learning using only unlabeled training data, which involves two key steps: 1) learning a good (e.g., compact and semantic) feature representation, and 2) modeling features of ID data without requiring class labels. For example, \cite{winkens2020contrastive} used contrastive training techniques SimCLR \cite{chen2020simple} to extract semantic features and proposed confusion log probability to determine whether a test example is a near or far OOD example. Using experiments, they show their approach is scalable to high-dimensional multimodal OOD examples. \cite{anonymous2021informative} also use contrastive loss based label-free training for self-supervised feature learning followed by OOD detection using Mahalanobis distance. 

Another line of label-free feature learning approaches for OOD detection uses flow-based generative models (e.g., VAEs, PixelCNNs, and Glow\cite{kingma2018glow}, allowing for the exact formulation of the marginal likelihood, to learn task-agnostic and semantic features to address the OOD detection problem. However, even sophisticated neural generative models trained to estimate feature density distribution (e.g., on CIFAR-10 images) can perform poorly on OOD detection, often assigning higher probabilities to OOD test examples than to IID test examples \cite{nalisnick2018deep}. Most recent research attempt to learn task-agnostic and semantic features for both IID and OOD images  \cite{zhang2020hybrid,shao2020open,nalisnick2019hybrid,chen2020informative}, yet unique challenges exist in learning task-agnostic and semantic representations. %Moreover the critical issue of error control in security-critical applications have not been sufficiently addressed.     

\subsection{Saliency guided training for enhancing DNN interpretability}
Saliency guided training has recently been shown to reduce noisy gradients used in predictions while retaining the predictive performance of the model. \cite{ismail2021improving} propose a saliency guided training by creating a new input by masking the features with low gradient values (salience) and encouraging the similarity between the new and original outputs. \cite{uddin2020saliencymix} develope a new approach that mixes the patches and labels using the salience density to select patches to dropout as a model regularization. \cite{chen2019robust} propose training objectives in classic robust optimization models to achieve robust Integrated Gradient (IG) attributions and demonstrate comparable prediction robustness (sometimes even better) while consistently improving attribution robustness. With these existing works, the generalizable features are expected to be learned via designing and optimizing a new salience-guided adversarial training objective as described below.

\section{Saliency Guided Adversarial Training}

Considering a classification problem on the input data $\{(X_i,y_i)\}_{i=1}^n$, a deep neural network model $f_\theta$ parameterized by $\theta$ is trained to predict the target $y$. The standard training involves minimizing the cross-entropy loss $\mathcal{L}$ over the training set as follows:
\begin{equation}
    \label{eq:celoss}
    \underset{\theta}{\mathrm{min}} \ \frac{1}{n} \sum_{i=1}^n \mathcal{L}(f_\theta(X_i),y_i).
\end{equation}
The model parameter $\theta$ is updated via one step of gradient descent with the learning rate $\alpha$:
\begin{equation}
    \label{eq:gd}
    \theta \leftarrow \theta - \alpha \cdot \frac{1}{m} \sum_{i=1}^m \nabla_\theta \mathcal{L}(f_\theta(X_i),y_i),
\end{equation}
on a mini-batch of $m$ samples $\{(X_i,y_i)\}_{i=1}^m$. We denote the gradient of the model output $f_\theta(X)$ with respect to the input $X$ as $\nabla_X f_\theta(X)$. 

Since the standard training procedure is based on ERM (expectation risk minimization) using stochastic gradient descent (SGD), the gradient of model w.r.t. the input (i.e., $\nabla_X f_\theta(X)$) may fluctuate sharply via small input perturbations \cite{smilkov2017smoothgrad}, e.g., adversarial noise. In this way, the model would probably learn some non-relevant features due to some uninformative local variations in partial derivatives. Furthermore, \cite{geirhos2020shortcut} observes that the traditional training approach introduces a nearly ideal scenario for DNN models to learn some spurious shortcut features, which do not exist in OOD test sets. 

Building on these intuitions, we propose saliency guided adversarial (SGA) training, a novel procedure to train the neural network models to learn the good features by suppressing non-relevant and eliminating shortcut features.

During saliency guided adversarial training, we augment the training set by generating a new training sample for each input sample $X$ by masking the features with low gradient values as follows:
\begin{equation}
    \label{eq:tildex}
    \tilde X = M_k(X,S(\nabla_X f_\theta(X))),
\end{equation}
where $S(\nabla)$ is a function that sorts the gradient of each feature from $X$ in the ascending sequence. $M_k(X,S(\nabla))$ is an input mask function, which replaces the $k$ lowest features from $X$ with random values within the feature range based on the order provided by $S(\nabla)$, as the non-relevant features usually have gradient values close to zero. 
$k$ is a tuning parameter, and its selection is based on the amount of nuisance information in a training sample. To further eliminate the shortcut features, we generate an adversarial example for the new sample $\tilde X$ as:
\begin{equation}
    \label{eq:xprime}
    X^\prime = \tilde X + \delta^\star,
\end{equation}
where $\delta^\star$ is estimated as:
\begin{equation}
    \label{eq:delta}
    \delta^\star = \underset{|\delta|_p \leq \epsilon} {\arg \max}\mathcal{L} (f_\theta(\tilde X + \delta),y),
\end{equation}
and $p$ can be $0,1,2, \ldots$ and $\infty$. In most cases, the perturbation budget $\epsilon$ is small so that the perturbations are imperceptible to human eyes. In our case, the adversarial example $X^\prime$ is generated based on the masked input $\tilde X$. Thus, $X^\prime$ does not contain the non-relevant features but has some shortcut features compared to the clean input $X$. 

$X^\prime$ is then passed through the model, resulting in an output $f_\theta(X^\prime)$. In addition to the classification loss used in the traditional training, saliency guided adversarial training adds another regularization term that minimizes the Kullback–Leibler (KL) divergence between $f_\theta(X)$ and $f_\theta(X^\prime)$. This regularization term ensures the model produces similar output probability distributions over labels for the original clean input $X$ and the masked adversarial example $X^\prime$. For this to happen, the model is ensured to learn the good features that ensure generalization performance on OOD test set.  

The optimization problem for our saliency guided adversarial training is:
% \begin{align}
%     \label{eq:loss}
%     \underset{\theta}{\mathrm{minimize}} \ \frac{1}{n} \sum_{i=1}^n \big[&\mathcal{L}(f_\theta(X_i),y_i) + \\
%     &\lambda D_{KL}(f_\theta(X_i)||f_\theta(X^\prime_i))\big],
% \end{align}
\begin{equation}
   \label{eq:loss}
    \underset{\theta}{\mathrm{min}} \frac{1}{n} \sum_{i=1}^n \big[\mathcal{L}(f_\theta(X_i),y_i) + \lambda D_{KL}(f_\theta(X_i)||f_\theta(X^\prime_i))\big],
\end{equation}
where $\lambda$ is a hyperparameter to leverage the importance of the cross-entropy classification loss and the KL divergence regularization term. Since this loss function is differentiable with respect to $\theta$, it can be optimized using existing gradient-based optimization methods. We show the saliency guided adversarial training procedure in Algorithm \ref{alg}. 

\begin{algorithm}
\caption{Saliency Guided Adversarial Training}\label{alg}
\begin{algorithmic} 
\STATE \textbf{Require:} Training Sample $X$, \# of features to be masked $k$, attack order $p$, perturbation budget $\epsilon$, learning rate $\tau$, hyperparameter $\lambda$, initialized model $f_\theta$
\FOR{epochs} 
    \FOR{minibaches} 
        \STATE \textbf{Create the masked input:}
            \STATE 1. Get sorted index $I$ for the gradient of output with respect to the input: $I = S(\nabla_X f_\theta(X))$
            \STATE 2. Mask bottom $k$ features of the original input: $\tilde X = M_k(X,I)$
        \STATE \textbf{Generate the adversarial example:}
            \STATE 1. Compute $\delta$: 
            $\delta^\star = \underset{|\delta|_p \leq \epsilon} {\arg \max}\mathcal{L} (f_\theta(\tilde X + \delta),y)$,
            \STATE 2. Generate the adversarial example: $X^\prime = \tilde X + \delta^\star$
        \STATE \textbf{Compute the loss:}
            \STATE $\mathcal{L}_i = \mathcal{L}(f_{\theta_i}(X),y) + \lambda D_{KL}(f_{\theta_i}(X)||f_{\theta_i}(X^\prime))$
       \STATE \textbf{Update $\theta$:}     
            \STATE $\theta_{i+1} = \theta_i - \tau \nabla_{\theta_i}\mathcal{L}_i$
    \ENDFOR
\ENDFOR 
\STATE \textbf{Return:} $f_\theta$
\end{algorithmic}
\end{algorithm}

\section{Experiments and Results}

In this section, we first define IID and OOD test sets in medical imaging domain and then explain how we built dataset for the COVID-19 detection task. Our proposed training approach is then evaluated qualitatively and quantitatively. Finally, the ablation analysis is performed to assess the effect of hyper-parameters.

\begin{figure*}[ht]
	\centering
	\includegraphics[width=0.99\textwidth]{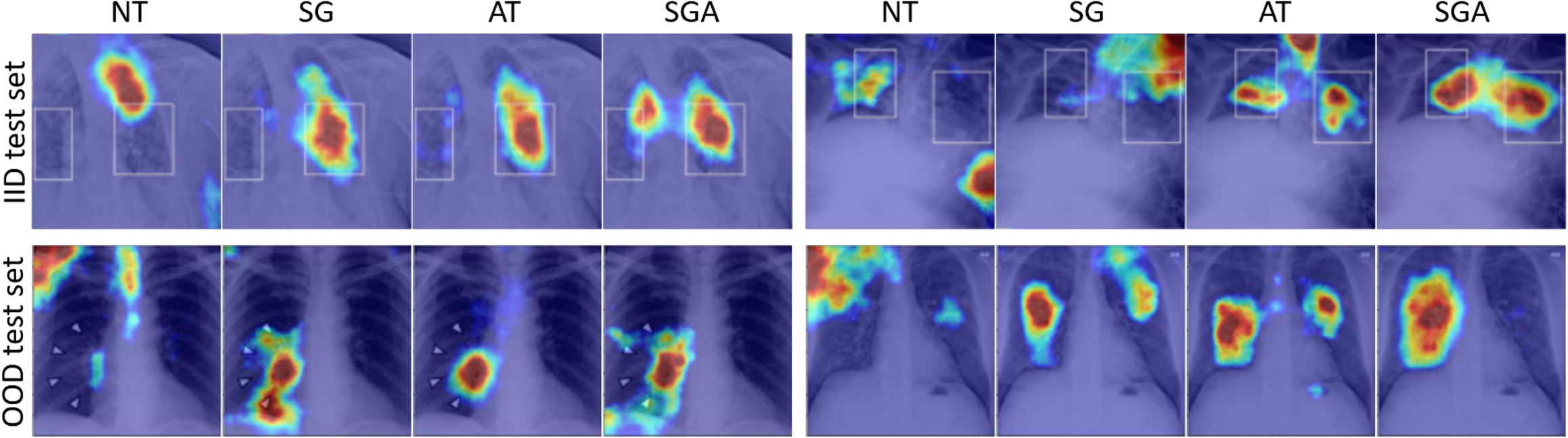}
	\caption{Examples to illustrate the features used by models to detect COVID-19+ in both IID and OOD test set. The bounding boxes in IID test set (top row) represents radiologists' annotated pathological features. Note we do not have similar annotations in OOD test set (bottom row). }
	\label{fig:qualitative}
\end{figure*}

%We use experiments to demonstrate that: 1): solve over-fitting feature 2) solve shortcut feature.

\paragraph{IID and OOD test sets} In machine learning, it's typical to randomly divide the available data into a training/validation and test set, with the former being used to select and teach the model to perform a particular task, and the latter being used to check the model's performance. One common assumption is that those two datasets are drawn from the same distribution. In relation to the training dataset, this test set is then referred as IID data \cite{geirhos2020shortcut}. Aside from the IID data, recent studies \cite{sehwag2019analyzing} evaluate the performance of AI systems on OOD data, which are systematically different from the IID data with a significant distribution shift. For example, in the medical domain, a test set acquired from different hospitals from the training dataset can be treated as OOD data \cite{degrave2021ai}. 

\paragraph{COVID-19 dataset} We select one benchmark data set and generate another dataset to evaluate our method for COVID-19 detection tasks. Dataset I is from SIIM-FISABIO-RSNA COVID-19 detection competition \cite{lakhani20212021}, which is used as the IID data set for training, validating, and internal testing. The dataset comprises 6334 CXR scans that are labeled by a panel of experienced radiologists with appearance and bounding box of COVID-19 opacities. The Dataset I is split into training, validation, and testing sets by a ratio of $6:2:2$. 

Dataset II is used for external test (OOD) only, which consists of COVID-19-positive X-ray in the GitHub-COVID repository \cite{cohen2020covid} collected from some public figures and other online sources with different geographic origins. Similar to \cite{li2020covid}, we further supplement these figures with COVID-19-negative (`No Findings') X-rays from the ChestX-ray14 dataset \cite{wang2017chestx}, which originate from a single hospital in the United States. It is important to note that the samples in Dataset I may contain COVID-19-negative CXRs from individuals with unknown pulmonary diseases, whereas the COVID-19-negative samples in Dataset II come from healthy individuals. As a result, the task of detecting positive COVID cases from the OOD test set can be less challenging than from the IID test set because the COVID negative cases in the former can be separated from COVID-19-positive cases more easily, giving rise to an enhanced performance (as opposed to degraded performance) in OOD test set using enhanced SGA training,  

\paragraph{Experiment settings} We use ResNet-18 \cite{he2016deep} pre-trained with ImageNet as the DNN architecture. which is trained with the SGD optimizer for 30 epochs with a batch size of 64. The adversarial samples $X^\prime$ are generated by FGSM for each minibatch with a uniformly sampling perturbation from the interval $[0.01,0.05]$ during the training process. The hyperparameter $k$ and $\lambda$ are fine-tuned as 0.1 and 1 respectively. The model that achieve the best performance on the validation set are used for IID and OOD testing. In order to demonstrate the effectiveness of our approach, we perform experiments comparing the performance of our SGA with three baseline methods: natural training (NT) with cross-entropy loss only, FGSM-based adversarial training (AT), and saliency guided training (SG). We show the heat map interpretations generated by FullGrad \cite{srinivas2019full}, which highlights the most salient regions of each CXR image that contribute mostly to the output, to illustrate the features exploited by the pre-trained models for COVID-19 detection.

\paragraph{Qualitative  evaluation} Figure \ref{fig:qualitative} shows the heat maps generated from the models trained with four competing training methods on two examples from IID and OOD test sets, respectively. NT generates the worst heat map interpretations on IID test set (top row) since the models seemly just learn some {\it non-relevant} features (e.g., those corresponding to the backbone shown in the left IID/NT panel) and/or some {\it shortcut} features (e.g., some special tags lying on the borders shown in the right IID/NT panel). This problem is aggravated on the OOD test set shown in the panels of the bottom row. Although SG and AT achieve slightly better interpretations than NT, SG still can not eliminate the {\it shortcut} features and AT seems to be plagued by some {\it non-relevant} features. On the contrary, the models trained with our SGA trend to use the COVID-19 pathological features (within the lungs inside the annotated bounding boxes) to detect COVID-19 in both IID and OOD test sets. This figure indicates that the regularization term added to our SGA training objective successfully learns good features by preventing the model from extracting {\it shortcut} and {\it non-relevant} features.

\begin{figure}[ht]
	\centering
	\includegraphics[width=0.45\textwidth]{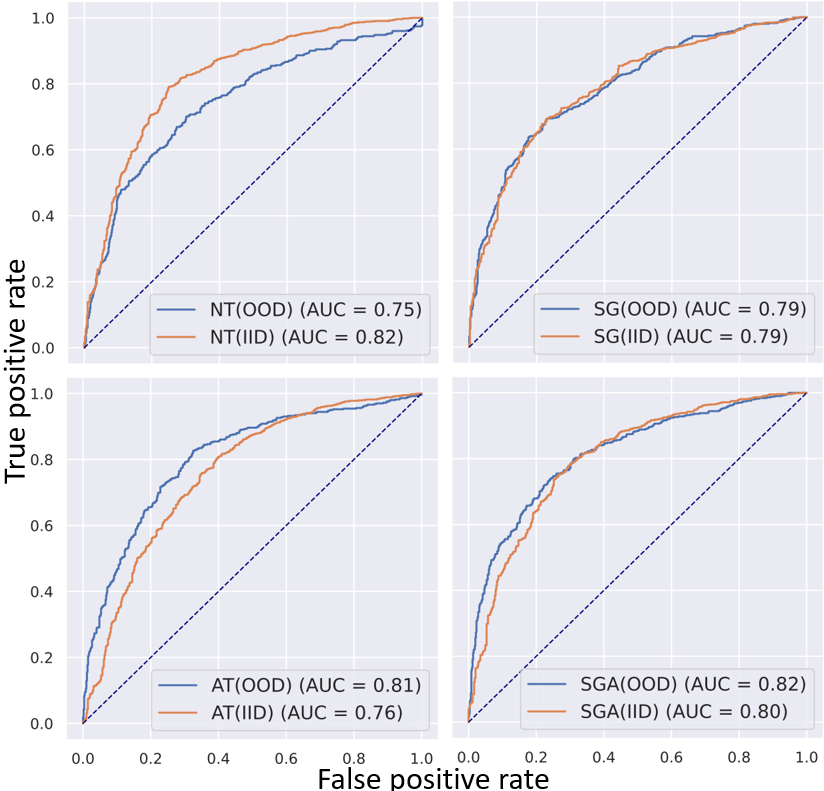}
	\caption{Model evaluations with receiver operating characteristic (ROC) curves, which show the performances on both internal test set (IID) and an external test set (OOD). The difference between IID and OOD test set performance is the performance degradation. Only the model trained with NT has a performance drop of $8.5\% (0.07/0.82)$ between IID and OOD test sets.}
	\label{fig:quantitative}
\end{figure}

\paragraph{Quantitative evaluation} In addition to the qualitative examples presented above, we also conduct quantitative experiments to validate our SGA method and compare with three baselines using the area under the curve (AUROC) for the imbalanced binary classification task. Figure \ref{fig:quantitative} demonstrates that SGA has the best average performance (0.81) on both IID and OOD test sets compared to other baselines: NT (0.78), SG (0.79), AT (0.79). A more important consideration is the performance drop from IID to OOD test set, which indicates whether the model uses {\it shortcut} and/or {\it non-relevant} features to make predictions. It is striking to note that the performances of AT and SGA increase from IID to OOD test set, which indicates (1) both training schemes learn and leverage good features and (2) the COVID prediction task on the OOD test set (Dataset II) is less challenging since the COVID-19-negative cases are free of lung diseases and thus comparatively more contrasting. The model trained with NT has a significant performance drop of $8.5\% (0.82 \rightarrow 0.75, 0.07/0.82)$, indicating that the model trained with NT uses shortcut and/or non-relevant features to make the prediction. This is consistent with what we demonstrate in the qualitative evaluation section.

\paragraph{Hyper-parameter $k$}

As previously indicated, assuming gradient-based explanation approaches interpret the model's predictions accurately, non-relevant features should have small gradient values. Based on this insight, we remove the $k$ lowest features from the input picture $X$ so as to encourage the model to learn good features. Note that the gradient value generated by Saliency Map can be negative. Higher negative values indicate that its absence contributes to an increased score of the class. These regions might be other objects in the background that cause the model to make incorrect predictions, and masking those region enables the model to concentrate on the foreground. As a result, in contrast to the original paper \cite{simonyan2013deep}, which used absolute gradient values to demonstrate the significance of features, we directly ordered the features and eliminated the lowest $k$ of features during training. 

The selection of $k$ depends on how much non-relevant information is in a training set. We chose small $k$ for our COVID-19 detection experiment because CXR images have little and clear backgrounds. Table \ref{table:k} show the result of ablation study on the hyper-parameter $k$, which is tuned from 0 to 0.2. Compared to AT, the model's performance increased by masking small amount of non-relevant features and achieve the best average performance on IID and OOD test sets when $k=0.1$.

\paragraph{Hyper-parameter $\lambda$} $\lambda$ is used to balance the contributions of NT and SGA regularization in the training objective. As shown in Table \ref{table:l}, compared to NT, the models trained with SGA have significantly better performance on OOD test set.

\begin{table}[]
\centering
\begin{tabular}{@{}c|ccccc@{}}
\toprule
$k$                & 0.00   & 0.05  & 0.10  & 0.15  & 0.20   \\ \midrule
IID Test         & 0.76  & 0.81 & 0.80 & 0.80 & 0.81  \\
OOD Test         & 0.81  & 0.78 & 0.82 & 0.79 & 0.80  \\ \midrule
Difference & +0.05 & -0.03 & +0.02 &  -0.01 & -0.01 \\
Average     & 0.79  & 0.80 & \textbf{0.81} & 0.79 & 0.80  \\ \bottomrule
\end{tabular}
	\caption{Ablation analysis of hyper-parameter $k$. Note that when $k = 0$, the model is trained by AT only.}
	\label{table:k}
\end{table}

% \usepackage{booktabs}
\begin{table}[]
\centering
\begin{tabular}{@{}c|ccccc@{}}
\toprule
$\lambda$   & 0      & 0.5 & 1     & 1.5 & 2    \\ \midrule
IID Test   & 0.82  & 0.78 & 0.80 & 0.78 & 0.78  \\
OOD Test   & 0.75  & 0.81 & 0.82 & 0.81 & 0.82  \\ \midrule
Difference & -0.07 & +0.03 & +0.02 & +0.03 & +0.04 \\
Average    & 0.79  & 0.79 & \textbf{0.81} & 0.79 & 0.80  \\ \bottomrule
\end{tabular}
	\caption{Ablation analysis of hyper-parameter $\lambda$. Note that when $\lambda = 0$, the model is trained by NT.}
	\label{table:l}
\end{table}

\section{Conclusions}

Existing DNN training methods can exploit non-relevant and shortcut features for prediction, which may account for the performance degradation on test set, particularly on OOD test sets. To overcome this limitation, we propose a novel saliency-guided adversarial training scheme for learning good features and empirically demonstrate its strong performance on CXR based OOD test sets, opening a new avenue for tackling the failure of medical imaging system in new hospitals or on new test sets.       

%In the future, XXX
% \begin{table*}[]
% \centering
% \begin{tabular}{@{}c|ccc|cc@{}}
% \toprule
% Dataset & \multicolumn{3}{c|}{IID dataset} & \multicolumn{2}{c}{OOD dataset} \\ \midrule
% Training & Accuracy & AUROC & Dice Score & Accuracy & AUROC \\ \midrule
% Baseline & 78.1\% & 0.821 & 0.70 & 64.9\% & 0.747 \\
% Saliency Guided & 75.3\% & 0.790 & X & 67.2\% & 0.790 \\
% Adversarial & 73.1\% & 0.778 & X & 75.3\% & X \\
% Our method & 77.2\% & 0.802 & 0.75 & 71.6\% & 0.817 \\ \bottomrule
% \end{tabular}
% \caption{Quantitative result on both IID and OOD datasets}\label{table1}

% \end{table*}

% \section*{Acknowledgements}

% \textbf{Do not} include acknowledgements in the initial version of
% the paper submitted for blind review.

% If a paper is accepted, the final camera-ready version can (and
% probably should) include acknowledgements. In this case, please
% place such acknowledgements in an unnumbered section at the
% end of the paper. Typically, this will include thanks to reviewers
% who gave useful comments, to colleagues who contributed to the ideas,
% and to funding agencies and corporate sponsors that provided financial

\newpage

\nocite{langley00}

\bibliography{SaTC}
\bibliographystyle{icml2021}

% \appendix
% \section{Do \emph{not} have an appendix here}

% \textbf{\emph{Do not put content after the references.}}
% %
% Put anything that you might normally include after the references in a separate
% supplementary file.

% We recommend that you build supplementary material in a separate document.
% If you must create one PDF and cut it up, please be careful to use a tool that
% doesn't alter the margins, and that doesn't aggressively rewrite the PDF file.
% pdftk usually works fine. 

% \textbf{Please do not use Apple's preview to cut off supplementary material.} In
% previous years it has altered margins, and created headaches at the camera-ready
% stage. 
%%%%%%%%%%%%%%%%%%%%%%%%%%%%%%%%%%%%%%%%%%%%%%%%%%%%%%%%%%%%%%%%%%%%%%%%%%%%%%%
%%%%%%%%%%%%%%%%%%%%%%%%%%%%%%%%%%%%%%%%%%%%%%%%%%%%%%%%%%%%%%%%%%%%%%%%%%%%%%%

\end{document}